\documentstyle[12pt]{article}

\begin{document}

\newcommand{\dfrac}[2]{\displaystyle{\frac{#1}{#2}}}

{\it University of Shizuoka}

\hspace*{9.5cm} {\bf Revised version of}\\[-.3in]

\hspace*{9.5cm} {\bf US-99-01}\\[-.3in]

\hspace*{9.5cm} {\bf May 1999}\\[.3in]

\vspace*{.4in}

\begin{center}

{\large\bf Universal Seesaw Mass Matrix Model\\
with an S$_3$ Symmetry} \\[.3in]

{\bf Yoshio Koide}\footnote{
E-mail: koide@u-shizuoka-ken.ac.jp} \\

Department of Physics, University of Shizuoka \\ 
52-1 Yada, Shizuoka 422-8526, Japan \\[.1in]

\vspace{.3in}

{\large\bf Abstract}\\[.1in]

\end{center}

\begin{quotation}
Stimulated by the phenomenological success of the universal
seesaw mass matrix model, where the mass terms for quarks and
leptons $f_i$ ($i=1,2,3$) and hypothetical super-heavy fermions
$F_i$ are given by $\overline{f}_L m_L F_R +\overline{F}_L m_R f_R
+ \overline{F}_L M_F F_R + h.c.$ and the form of $M_F$ is
democratic on the bases on which $m_L$ and $m_R$ are diagonal,
the following model is discussed: 
The mass terms $M_F$  are invariant under the permutation symmetry 
S$_3$,  and the mass terms $m_L$ and $m_R$ are generated by  
breaking the S$_3$ symmetry spontaneously.
The model leads to an interesting relation for the charged lepton 
masses.
\end{quotation}

\vfill
\newpage


The universal seesaw mass matrix model\cite{USMM} is 
one of the most promising candidates of unified quark and lepton 
mass matrix models. 
The model has hypothetical fermions
$F_i$ ($F=U,D,N,E$; $i=1,2,3$) in addition to the conventional 
quarks and leptons $f_i$ ($f=u, d, \nu, e$; $i=1,2,3$), 
and these fermions are assigned to $f_L = (2,1)$, 
$f_R = (1,2)$, $F_L = (1,1)$ and $F_R = (1,1)$ of 
SU(2)$_L \times$ SU(2)$_R$.
The 6 $\times$ 6 mass matrix which is sandwiched
between the fields ($\overline{f}_L, \overline{F}_L$)
and ($f_R, F_R$) is given by
$$
M^{6 \times 6} =
\left( \begin{array}{cc}
0 & m_L\\
m_R & M_F
\end{array} \right) \ ,
\eqno(1)
$$
where $m_L$ and $m_R$ are universal for all fermion
sectors ($f=u, d, \nu, e$) and only $M_F$ have
structures dependent on the flavors $F$.
For $\Lambda_L < \Lambda_R \ll \Lambda_S$, where 
$\Lambda_L = O(m_L), \Lambda_R = O(m_R)$ and
$\Lambda_S = O(M_F)$, the $3\times 3$ mass matrix
$M_f$ for the fermions $f$ is given by the
well-known seesaw expression
$$
M_f \simeq - m_L M^{-1}_F m_R \ .
\eqno(2)
$$
Thus, the model answers the question why the masses of 
quarks (except for top quark) and charged leptons are
so small compared with the electroweak scale
$\Lambda_L$ ($\sim$ 10$^2$ GeV).
On the other hand, in order to understand the observed
fact $m_t \sim \Lambda_L$, we put the ansatz
\cite{KFzp,Morozumi} $ {\rm det} M_F = 0$
for the up-quark sector ($F=U$).
Then,  one of the fermion
masses $m(U_i)$ is zero [say, $m(U_3)=0$],
so that the seesaw mechanism does not work for
the third family, i.e., the fermions ($u_{3L}, U_{3R}$)
and ($u_{3R}, U_{3L}$) acquire masses of $O(m_L)$ and
$O(m_R)$, respectively. 
We identify $(u_{3L},U_{3_R})$ as the top quark 
$(t_L, t_R)$.
Thus,  we can understand the question
why only the top quark has a mass of the order of
$\Lambda_L$.

For the numerical results, excellent agreements
with the observed values of the quark masses and
Cabibbo-Kobayashi-Maskawa \cite{CKM} (CKM) matrix are
obtained by putting the following assumptions $\cite{KFzp}$: 

\noindent
(i) The mass matrices $m_L$ and $m_R$ have the same structure
$$
m_R = \kappa m_L \equiv m_0 \kappa Z \ .
\eqno(3)
$$
(ii) The mass matrix $M_F$ is given by the form
$$
M_F = m_0 \lambda ({\bf 1} + 3b_f X),
\eqno(4)
$$
$$
{\bf 1} = 
\left( \begin{array}{ccc}
1 & 0 & 0 \\
0 & 1 & 0 \\
0 & 0 & 1 \\
\end{array} \right)
,\ \ \ 
X = \frac{1}{3}
\left( \begin{array}{ccc}
1 & 1 & 1 \\
1 & 1 & 1 \\
1 & 1 & 1 \\
\end{array} \right) ,
\eqno(5)
$$
on the basis on which the matrix $Z$ is diagonal, i.e.,
$$
Z = {\rm diag} (z_1, z_2, z_3),
\eqno(6)
$$
where $z_1^2 + z_2^2 + z_3^2 = 1$. 

\noindent
(iii) The parameter $b_f$ for the charged lepton sector is
given by $b_e$ = 0, so that in the limit of $\kappa/\lambda
\ll 1$, the parameters $z_i$ are given by 
$$
\frac{z_1}{\sqrt{m_e}} = \frac{z_2}{\sqrt{m_\mu}} =
\frac{z_3}{\sqrt{m_\tau}} = \frac{1}{\sqrt{m_e + m_\mu + m_\tau}}
\eqno(7)
$$
(iv) Then, the up- and down-quark masses are successfully
given by the choice of $b_u = -1/3$ and $b_d = -e^{i \beta d}$
($\beta_d = 18^{\circ}$), respectively. The CKM matrix is also
successfully obtained.

In this phenomenological success, the assumption that the mass
matrix $M_F$ is the democratic type is essential. The form
of $M_F$, (4), is invariant under the permutation symmetry\cite{S3}
S$_3$ for ($F_1, F_2, F_3$), 
while the form of $m_L$ ($m_R$) is
not invariant under the permutation symmetry
S$_3$ for $(F_1, F_2, F_3)$ and $(f_1, f_2, f_3)$.
In this paper, we consider that the mass terms $m_L$ ($m_R$)
are generated by breaking the S$_3$ symmetry not explicitly,
but spontaneously at $\mu=\Lambda_L$ ($\mu=\Lambda_R$).
For this purpose, we introduce three SU(2)$_L$-doublet
Higgs scalars $(\phi_{1L}, \phi_{2L}, \phi_{3L})$,
which obey to the permutation symmetry S$_3$ as well as
$(F_1, F_2, F_3)$ and $(f_1, f_2, f_3)$.
(We also assume three SU(2)$_R$-doublet Higgs scalars.)
The purpose of the present paper is to discuss the 
possible structure of $m_L$ ($m_R$) under this S$_3$ symmetry.

The Yukawa interactions which generate the mass matrix
$m_L$ are given by
$$
y_L \sum_i [(\overline{\nu}_{iL} \  \overline{e}_{iL})
\left[
\left( \begin{array}{c}
\phi^+_{iL} \\
\phi^0_{iL} \\
\end{array} \right)
E_{iR} + 
\left( \begin{array}{c}
\overline{\phi}^0_{iL} \\
-\phi^-_{iL} \\
\end{array} \right)
N_{iR} 
\right]  + 
({\rm quark\ sectors}) \ ,
\eqno(8)
$$
Hereafter, for convenience, we drop the index L.
The most simple form of the S$_3$ invariant potential of the
Higgs scalars ($\phi_1, \phi_2, \phi_3$) is
$$
V_1 = \mu^2 \sum_i (\overline{\phi}_i \phi_i) + \frac{1}{2} \lambda_1
[\sum_i (\overline{\phi}_i \phi_i)]^2,
\eqno(9)
$$
where $(\overline{\phi}_i \phi_i) = \phi^-_i \phi^+_i + 
\overline{\phi}^0_i \phi^0_i$. Note that the term
$$
V_2 = \eta_1 (\overline{\phi}_{\sigma} \phi_{\sigma})
(\overline{\phi}_{\pi} \phi_{\pi} + \overline{\phi}_{\eta} \phi_{\eta}),
\eqno(10)
$$
is also S$_3$-invariant, where
$$
\phi_{\pi} = \frac{1}{\sqrt{2}}(\phi_1 - \phi_2),
\eqno(11)
$$
$$
\phi_{\eta} = \frac{1}{\sqrt{6}}(\phi_1 + \phi_2 - 2\phi_3),
\eqno(12)
$$
$$
\phi_{\sigma} = \frac{1}{\sqrt{3}}(\phi_1 + \phi_2 + \phi_3),
\eqno(13)
$$
and
$$
\sum_i (\overline{\phi}_i \phi_i) = (\overline{\phi}_{\pi} \phi_{\pi})
+ (\overline{\phi}_{\eta} \phi_{\eta})
+ (\overline{\phi}_{\sigma} \phi_{\sigma}).
\eqno(14)
$$

We assume that the potential of the Higgs scalars
($\phi_1, \phi_2, \phi_3$) is given by
$$
V = V_1 + V_2.
\eqno(15)
$$
Then, the conditions for the vacuum expectation values
$v_i \equiv \langle \phi^0_i \rangle$ at which the potential (15)
takes the minimum are
$$
\mu^2 + \lambda_1 \sum_i |v_i|^2 
+ \eta_1(|v_{\pi}|^2 + |v_{\eta}|^2) = 0,
\eqno(16)
$$
$$
\mu^2 + \lambda_1 \sum_i |v_i|^2 + \eta_1 |v_{\sigma}|^2 = 0,
\eqno(17)
$$
so that
$$
|v_{\sigma}|^2 = |v_{\pi}|^2 + |v_{\eta}|^2
= \frac{-\mu^2}{2 \lambda_1 + \eta_1},
\eqno(18)
$$
From the relations (13), (14) and (18), we obtain
$$
|v_1|^2 + |v_2|^2 + |v_3|^3 = 2 |v_{\sigma}|^2
= \frac{2}{3} |v_1 + v_2 + v_3|^2,
\eqno(19)
$$
which means the relation\cite{Koide-me}
$$
m_e + m_{\mu} + m_{\tau}
= \frac{2}{3}(\sqrt{m_e} + \sqrt{m_{\mu}} + \sqrt{m_{\tau}})^2,
\eqno(20)
$$
from the relation (7). The relation (20) is excellently satisfied
by the observed values of the charged lepton masses, i.e., the
observed values of $m_e$ and $m_{\mu}$ give the predicted
value $m_{\tau} = 1776.97$ MeV in agreement with the observed
value \cite{pdg} $m^{exp}_{\tau} = 1777.05^{+0.29}_{-0.26}$ MeV.
[We should not take this excellent agreement too rigidly,
because the electromagnetic corrections to the observed values
spoil the agreement of $m_\tau(\mu)$, for example, to
1.2\% at the energy scale $\mu=m_Z=91.2$ GeV.
However, note that the relation (7) is an approximate one.
When we define $m_L=m_0 Z_L$ and $m_R=m_0\kappa Z_R$, the 
values of $Z_L(\mu)$ and $Z_R(\mu)$ are dependent of on the
energy scale $\mu$, so that the relation $Z_L(\mu)=Z_R(\mu)$
is an approximate relation even if it is exact at a unification 
energy scale $\mu=\Lambda_X$.
In order to examine the validity of the relation (20), we 
must know the energy scale structures in the seesaw model
(e.g., the energy scales of $m_R$, $M_F$, and so on).
At present, we consider that the relation (19) is sill
worth noting.]

Explicitly, from the relations (11) - (13), the charged lepton
masses $m^e_i$ are given by
$$
\sqrt{m_{\tau}} = \sqrt{m^e_1} \propto v_1 =
\left(\frac{1}{\sqrt{2}} \cos\theta + \frac{1}{\sqrt{6}} \sin\theta
+ \frac{1}{\sqrt{3}}\right)v_\sigma \ ,
\eqno(21)
$$
$$
\sqrt{m_\mu} = \sqrt{m^e_2} \propto v_2 =
\left(-\frac{1}{\sqrt{2}} \cos\theta + \frac{1}{\sqrt{6}} \sin\theta
+ \frac{1}{\sqrt{3}}\right)v_\sigma \ ,
\eqno(22)
$$
$$
\sqrt{m_e} = \sqrt{m^e_3} \propto v_3 =
\left(-\sqrt{\frac{2}{3}} \sin\theta + \frac{1}{\sqrt{3}}\right)v_\sigma 
\ ,
\eqno(23)
$$
where
$$
v_\pi = v_\sigma \cos\theta \ ,\ \ \ v_\eta = v_\sigma \sin\theta \ ,
\eqno(24)
$$
Since the model is $\phi_\pi \leftrightarrow \phi_\eta$ symmetric,
it is likely that the vacuum expectation values satisfy the relation
$v_\pi \simeq v_\eta$, i.e., $\sin\theta \simeq \cos\theta \simeq
1/\sqrt{2}$. In the limit of $\sin\theta = \cos\theta = 1/\sqrt{2}$,
the electron mass becomes exactly zero. In order to give
$v_\pi \neq v_\eta$, we must add a small additional term to the
Higgs potential (15).
However, for a time, we will not touch the origin of
$m_e \neq 0$.

The potential (15) is not general form which is invariant under
the S$_3$ symmetry. 
The general S$_3$-invariant potential is given as a function of
$\overline{\phi}_{\sigma\alpha} \phi_{\sigma\beta}$ and 
$\overline{\phi}_{\pi\alpha} \phi_{\pi\beta}
+ \overline{\phi}_{\eta\alpha} \phi_{\eta\beta}$ (and also
${\phi}_{\sigma\alpha} \phi_{\sigma\beta}$ and 
${\phi}_{\pi\alpha} \phi_{\pi\beta}+{\phi}_{\eta\alpha} 
\phi_{\eta\beta}$, where $\alpha$ and $\beta$ are SU(2) indices.
For example, the potential
$$
V=\mu^2 \left[ (\overline{\phi}_{\sigma} \phi_{\sigma}) 
+k (\overline{\phi}_{\pi} \phi_{\pi}+
\overline{\phi}_{\eta} \phi_{\eta})\right]
$$
$$
+\frac{1}{2}\lambda \left[ 
a (\overline{\phi}_{\sigma} \phi_{\sigma})^2 
+b (\overline{\phi}_{\sigma} \phi_{\sigma})
 (\overline{\phi}_{\pi} \phi_{\pi}+
\overline{\phi}_{\eta} \phi_{\eta})
+c (\overline{\phi}_{\sigma} \phi_{\sigma})
 (\overline{\phi}_{\pi} \phi_{\pi}+
\overline{\phi}_{\eta} \phi_{\eta})^2
\right]\ ,
\eqno(25)
$$
is S$_3$-invariant, while the potential (25) with $k\neq 1$
and $a\neq c$ cannot give the relation (19).
In order to give the relation (19), the following condition
is required:
The potential is invariant under the exchange
$$
\overline{\phi}_{\sigma\alpha} \phi_{\sigma\beta} \leftrightarrow 
\overline{\phi}_{\pi\alpha} \phi_{\pi\beta}
+ \overline{\phi}_{\eta\alpha} \phi_{\eta\beta} \ ,
$$
$$
{\phi}_{\sigma\alpha} \phi_{\sigma\beta} \leftrightarrow 
{\phi}_{\pi\alpha} \phi_{\pi\beta}+{\phi}_{\eta\alpha} 
\phi_{\eta\beta} \ , \eqno(26)
$$
$$
\overline{\phi}_{\sigma\alpha} \overline{\phi}_{\sigma\beta} 
\leftrightarrow 
\overline{\phi}_{\pi\alpha} \overline{\phi}_{\pi\beta}
+ \overline{\phi}_{\eta\alpha} \overline{\phi}_{\eta\beta} \ .
$$
The most general form which is invariant under the exchange (26)
is given by $V=V_1+V_2+V_3$, where $V_3$ is given by
$$
V_3=\frac{1}{2}\lambda_2 \sum_i \sum_j (\overline{\phi}_i \phi_j)
(\overline{\phi}_j \phi_i)
+ \frac{1}{2}\lambda_3 \sum_i \sum_j (\overline{\phi}_i \phi_j)
(\overline{\phi}_i \phi_j) 
$$
$$
+\eta_2 \left[(\overline{\phi}_\sigma \phi_\pi)
(\overline{\phi}_\pi \phi_\sigma)
+(\overline{\phi}_\sigma \phi_\eta)
(\overline{\phi}_\eta \phi_\sigma)\right]
$$
$$
+\eta_3 \left[(\overline{\phi}_\sigma \phi_\pi)
(\overline{\phi}_\sigma \phi_\pi)
+(\overline{\phi}_\sigma \phi_\eta)
(\overline{\phi}_\sigma \phi_\eta) +h.c. \right]
\ .
\eqno(27)
$$
Then, the potential $V$ leads to the relation
$$
|v_\sigma|^2 = |v_\pi|^2 + |v_\eta|^2
= \frac{-\mu^2}{2(\lambda_1+\lambda_2+\lambda_3)
+\eta_1+\eta_2+2\eta_3}\ ,
\eqno(28)
$$
instead of (18), so  that we can again obtain the relation (20).

In Table I, we give the masses of the physical Higgs bosons
$H_S^0$, $H_A^0$, $H_B^0$, $\chi_A^0$, $\chi_B^0$, $\chi_A^\pm$, 
and $\chi_B^\pm$, which are defined by
$$
\phi_i \equiv
\left( \begin{array}{c}
\phi_i^+ \\
\phi_i^0 \\
\end{array} \right)
= \frac{1}{\sqrt{2}}
\left( \begin{array}{c}
i\sqrt{2} \chi_i^+ \\
H_i^0 - i \chi_i^0 \\
\end{array} \right)
,
\eqno(29)
$$
$$
\left( \begin{array}{c}
\phi_S \\
\phi_A \\
\phi_B \\
\end{array} \right)
= \frac{1}{v_0} \left( \begin{array}{ccc}
v_1 & v_2 & v_3 \\
v_1 -\sqrt{\frac{2}{3}} v_0 & v_2 -\sqrt{\frac{2}{3}} v_0
& v_3 -\sqrt{\frac{2}{3}} v_0 \\
\sqrt{\frac{2}{3}} (v_3-v_2) & \sqrt{\frac{2}{3}} (v_1-v_2)
& \sqrt{\frac{2}{3}} (v_2-v_3) \\
\end{array} \right)
\left( \begin{array}{c}
\phi_1 \\
\phi_2 \\
\phi_3 \\
\end{array} \right)
$$
$$
= \frac{1}{v_0} \left( \begin{array}{ccc}
v_\pi & v_\eta & v_\sigma \\
v_\pi & v_\eta & -v_\sigma \\
\sqrt{2} v_\eta & -\sqrt{2} v_\pi & 0 \\
\end{array} \right)
\left( \begin{array}{c}
\phi_\pi \\
\phi_\eta \\
\phi_\sigma \\
\end{array} \right) \ ,
\eqno(30)
$$
$$
v_0^2 = v_1^2 + v_2^2 + v_3^2 = v_\pi^2 + v_\eta^2 + v_\sigma^2
= 2 v_\sigma^2\ .
\eqno(31)
$$
[The evaluations are analogous to those in Ref.\cite{nonet},
where the U(3)-family nonet Higgs scalars $\phi_i^j$ 
($i,j=1,2,3$) were assumed. We can read $\phi_i^j$ in 
Ref.\cite{nonet} as $\phi_i^i \rightarrow \phi_i$ ($i=1,2,3$) 
and $\phi_i^j \rightarrow 0$ ($i \neq j$).]
The Higgs components $\chi_S^\pm$ and $\chi_S^0$ are eaten by the
weak bosons $W^\pm$ and $Z$, respectively. The Higgs boson $H_S$
corresponds to that in the standard one Higgs boson model.
Note that the Higgs scalar $H_B$ is massless.
Also, $H_A$ is massless if the $\eta$-terms are absent, and
$\chi_A^\pm$, $\chi_B^\pm$, $\chi_A^0$, and $\chi_B^0$ are 
massless if the terms $V_3$ are absent.

In the present model, the flavor-changing neutral currents
(FCNC) effects do not appear in the charged lepton sector,
because the mass matrix of the charged leptons is diagonal. 
However, in the neutrino and quark sectors, the FCNC effects
appear through the exchanges of the neutral Higgs bosons
$H_A^0$, $H_B^0$, $\chi_A^0$, and $\chi_B^0$.
Although the FCNC in the neutrino sectors have a possibility
\cite{nu-FCNC} that they can offer an alternative 
mechanism to the neutrino oscillation hypothesis, they, in
general, bring unwelcome effects, especially, in the quark
sectors.
In order to avoid this problem, for example, we
must distinguish the Higgs scalars $\phi_i^u$ which couple to
the up-fermion sectors, from the scalars $\phi_i^d$ which
couple to the down-fermion sectors.
At present, this is an open question.

In conclusion, stimulated by the phenomenological success
of the universal seesaw mass matrix model \cite{KF-zp},
we have proposed a Higgs potential which is
invariant under the permutation symmetry S$_3$ for
($f_1,f_2,f_3$), ($F_1,F_2,F_3$) and ($\phi_1,\phi_2,\phi_3$),
and which leads to the relation (20) for the charged lepton
masses.
It is worth while to notice the model because of the 
agreement of the relation (20) with experiments, 
although it has a trouble in FCNC.

\vspace*{.2in}

\centerline{\Large\bf Acknowledgments}

The author would like to thank T.~Maskawa, J.~Arafune, I.~Sogami,
T.~Fukuyama and T.~Kurimoto 
for  helpful discussions and their useful comments.
This work was supported by the Grand-in-Aid for Scientific
Research, Ministry of Education, Science and Culture,
Japan (No.~08640386).

\vglue.2in


\newpage

{\bf Table I.}\ Physical Higgs boson masses,
where $v_0^2=v_1^2+v_2^2+v_3^2=(174\ {\rm GeV})^2$.
$$
\begin{tabular}{|c|c|c|c|} \hline
$\phi$ & $\chi^\pm$ & $\chi^0$ & $H^0$ \\ \hline
$m^2(\phi_S)$ &  eaten by $W^\pm$ &  eaten by $Z$
& $[2(\lambda_1+\lambda_2+\lambda_3)+\eta_1+\eta_2+2\eta_3]v_0^2$ \\
$m^2(\phi_A)$ & $-(\lambda_2+\lambda_3+\eta_2+2\eta_3)v_0^2$
& $-2(\lambda_3+2\eta_3)v_0^2$ & $-(\eta_1+\eta_2+2\eta_3)v_0^2$ \\
$m^2(\phi_B)$ & $-(\lambda_2+\lambda_3+\frac{1}{2}\eta_2+\eta_3)v_0^2$
& $-2(\lambda_3+\eta_3)v_0^2$ & $0$ \\ \hline
\end{tabular}
$$

\end{document}